# Substituent Effects on the Spin-Transition Temperature in Complexes with Tris(pyrazolyl) Ligands

**Hauke Paulsen**[1,*], **Lars Duelund**[2], **Axel Zimmermann**[2], **Frédéric Averseng**[1], **Michael Gerdan**[1], **Heiner Winkler**[1], **Hans Toftlund**[2], and **Alfred X. Trautwein**[1]

[1] Institut für Physik, Universität zu Lübeck, D-23538 Lübeck, Germany
[2] MEMPHYS – Center for Biomembrane Physics, Department of Chemistry, University of Southern Denmark, DK-5230 Odense, Denmark
[2] Department of Chemistry, University of Southern Denmark, 5230 Odense, Denmark

**Summary.** Iron (II) complexes with substituted tris(pyrazolyl) ligands, which exhibit a thermally driven transition from a low-spin state at low temperatures to a high-spin state at elevated temperatures, have been studied by Mössbauer spectroscopy and magnetic susceptibility measurements. From the observed spectra the molar high-spin fraction and the transition temperature have been extracted. All substituents, except for bromine, lead to a decrease of the transition temperature. Density functional calculations have been carried out to compare the experimentally observed shifts of the transition temperature with those derived from theory.

## Introduction

Spin-crossover complexes exhibit a transition from a low-spin (LS) to a high-spin (HS) state that can be reversibly induced by changing temperature or pressure or by irradiation with light [1]. These complexes are therefore very promising materials for display and memory devices. Gradual transitions, where the molar HS fraction $\gamma_{HS}(p,T)$ at given pressure $p$ changes smoothly over a large interval of temperature $T$, can be explained with a simple model [2], that explicitly neglects cooperativity:

$$\gamma_{HS}(p,T) = 1 / [1+\exp(\Delta G/k_B T)]. \qquad (1)$$

Here $\Delta G$ denotes the difference (HS-LS) of the Gibbs free energy,

$$G = E_{el} + E_{vib} + pV - TS, \qquad (2)$$

depending on the total electronic energy $E_{el}$, the vibrational energy $E_{vib}$, pressure $p$, volume $V$, temperature $T$, entropy $S$, and, implicitly, on $\gamma_{HS}$. This means that eq. (1) has to be solved iteratively. The molar HS fraction $\gamma_{HS}$ is used for the determination of many properties of spin-crossover materials, for instance the transition temperature $T_{1/2}$, which is defined by

$$\gamma_{HS}(p,T_{1/2}) = 1/2. \qquad (3)$$

The purpose of the present study is the investigation of the influence of substituents on the spin-transition temperature of iron(II) complexes with tris(pyrazolyl) ligands. The starting point for this investigation is a complex formed by an iron(II) center and two tris(pyrazol-1-yl)methane ligands (compound **1**[a], Figure 1) [3]. A variety of ligands can be obtained by substituting the hydrogens of the pyrazol rings (Scheme 1). In the present study four different ligands have been used: tris(3-methylpyrazol-1-yl)methane (compounds **2**[a], **2**[b], and **2**[c]), tris(4-methyl-pyrazol-1-yl)methane (compound **3**[a]), tris(4-

---

✱ Corresponding author. E-mail: paulsen@physik.uni-luebeck.de

bromo-pyrazol-1-yl)methane (compound **4**[b]), and tris(3,5-dimethylpyrazol-1-yl)methane (compound **5**[b]). The superscripts denote the counteranions $PF_6^-$ (a), $ClO_4^-$ (b), and $BF_4^-$ (c). One purpose of such substitutions is to shift $T_{1/2}$. A general understanding of such substituent effects could help to rationalize the design of spin-crossover complexes with certain properties which are useful for technical applications (e.g. $T_{1/2}$ at ambient temperature and wide hysteresis of the temperature-dependent molar HS fraction).

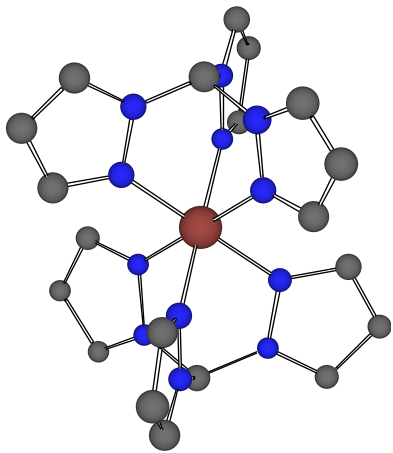

**Figure 1.** Molecular structure of the HS isomer of complex **1** derived from geometry optimization. Hydrogen atoms have been omitted for clarity.

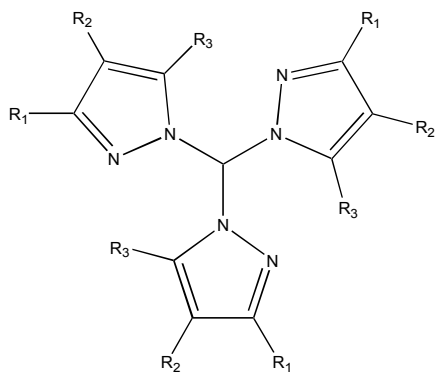

**Scheme 1.** Drawing of the ligands tris(pyrazol-1-yl)methane ($R_1,R_2,R_3$=H), tris(3-methylpyrazol-1-yl)methane ($R_1$=$CH_3$, $R_2,R_3$=H), tris(4-methyl-pyrazol-1-yl)methane ($R_1,R_3$=H, $R_2$=$CH_3$), tris(4-bromo-pyrazol-1-yl)methane ($R_1,R_3$=H, $R_2$=Br), and tris(3,5-dimethylpyrazol-1-yl)methane ($R_1,R_3$=$CH_3$, $R_2$=H) used for complexes **1**, **2**, **3**, **4**, and **5**, respectively.

Mössbauer transmission experiments and magnetic susceptibility measurements have been performed at various temperatures in order to find out whether these complexes exhibit spin-crossover behaviour and to determine $T_{1/2}$. At the same time electronic structure calculations have been performed to estimate $T_{1/2}$ from theory. These calculations have been carried out for isolated complexes *in vacuo*. This approximation neglects interactions between neighbouring complexes and interactions between the complexes and their counterions. Both interactions are known to considerably influence $T_{1/2}$. Calculations which do not regard these interactions can therefore at best be in qualitative agreement with the experiment. Nevertheless, such calculations for isolated complexes *in vacuo* may reveal information about the molecular contribution to substituent-induced shifts of $T_{1/2}$. This information can hardly be gained experimentally since any experiment with a solid sample will only reflect the combined influence of intra- and intermolecular interactions. Calculation of the normal modes of vibrations applying density functional theory (DFT) [4,5] have demonstrated that the molecular approximation can be useful also for spin-crossover complexes in the solid state. In future attempts we also plan to include intramolecular interactions in our calculations.

**Results and Discussion**

*Mössbauer Spectra*

The measured Mössbauer spectra for compounds **2**[a], **2**[b], **2**[c], **3**[a], **4**[b], and **5**[b] have been fitted using subspectra for the HS and the LS isomers (Figure 2), Mössbauer spectra for compound **1**[a] have been published earlier [3]. For the low-spin isomers at low temperature the measured quadrupole splittings are in a narrow range between 0.3 and 0.4 mm/s, the observed isomer shifts are in the range between 0.47 and 0.53 mm/s (Table 1). The influence of the substituents on the Mössbauer parameters is weak, and no significant influence of the counterions could be observed for compounds **2**[a], **2**[b], and **2**[c]. The calculated quadrupole splittings for the LS isomers range from 0.01 to 0.12 mm/s. This is most probably due the fact that the calculations have been performed for molecules *in vacuo* with $D_{3d}$ symmetry. In the solid sample small distortions of the $D_{3d}$ symmetry are expected.

**Table 1.** Measured quadrupole splitting $\Delta E_Q$ and isomer shift $\delta$ in mm/s at low temperatures (calculated values for $\Delta E_Q$ using the BLYP//6-311G method are given in brackets).

| Complex | $\Delta E_Q$ | | $\delta$ | |
|---|---|---|---|---|
| | LS | HS | LS | HS |
| **1** | 0.30 (0.10) | 2.20[d] (3.81) | 0.47 | 0.85[d] |
| **2**[a] | 0.39 (0.01) | 3.96 (3.75) | 0.53 | 1.13 |
| **2**[b] | 0.43 (0.01) | 3.71 (3.75) | 0.53 | 1.24 |
| **2**[c] | 0.36 (0.01) | 3.99 (3.75) | 0.53 | 1.13 |
| **3**[a] | 0.32 (0.09) | 3.55 ( — ) | 0.51 | 0.97 |
| **4**[b] | 0.35 (0.12) | 3.12 ((3.34)) | 0.48 | 1,41 |
| **5**[b] | — ( — ) | 3.99 ( — ) | — | 1.15 |

[a]with counterion $PF_6^-$, [b]with counterion $ClO_4^-$, [c]with counterion $BF_4^-$, [d]measured at room temperature

The Mössbauer parameters for the HS isomer are less straightforward to compare. At low temperatures usually no or only a small fraction of HS isomers are present leading to large error margins for the measured Mössbauer parameters. At higher temperatures, where a large HS fraction is present, a temperature-dependent reduction of $\Delta E_Q$ may be observed. This is due to thermal population of low-lying excited electronic states. The observed quadrupole splittings of HS isomers range between 3 and 4 mm/s, except for compound **1**[a], where a quadrupole splitting of 2.2 mm/s is reported [3]. However, due to the high transition temperature of about 355 K the low-temperature value is experimentally very difficult to access. The calculated quadrupole splitting of 3.81 mm/s suggests that the low-temperature quadrupole splitting of compound **1**[a] is quite similar to the corresponding values of the other compounds. For the isomer shift values between 0.9 and 1.4 mm/s have been observed. The calculated quadrupole splittings for the HS isomers are in quantitative agreement with the measured values, considering the uncertainty of the influence of spin-orbit coupling and thermal population of excited states.

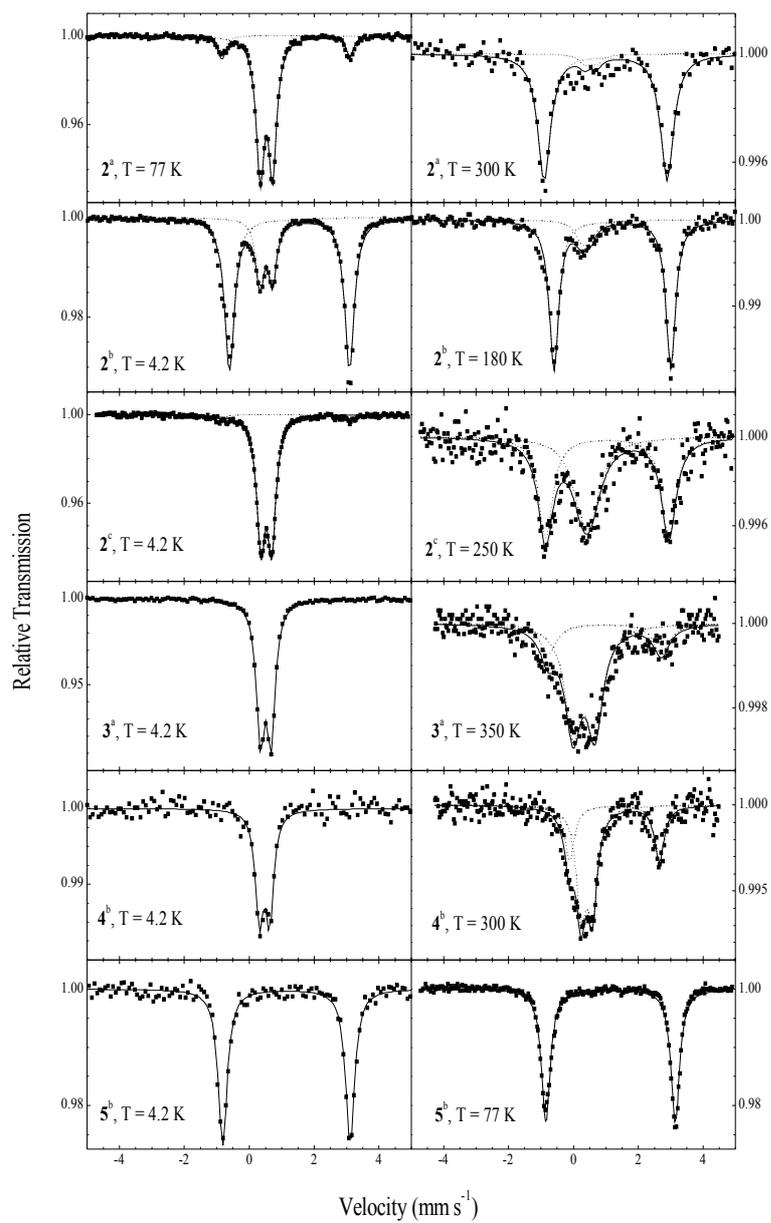

**Figure 2.** Mössbauer spectra of different complexes at temperatures as indicated.

The molar HS fraction $\gamma_{HS}$ can be retrieved in principle from the relation of the area of the HS subspectrum to the area of the LS subspectrum (Figure 2). For an accurate determination of $\gamma_{HS}$ the measured areas have to be corrected for the Lamb-Mössbauer factors which can be quite different for the HS and the LS isomers. Even if a small fraction of spin-crossover complexes is randomly distributed in a host lattice of other complexes, the molecular contribution to the Lamb-Mössbauer factor of the LS isomer can be different from the respective HS value [4,6].

*Magnetic Susceptibility*

The second experimental technique that has been applied here to monitor the temperature-dependent molar HS fraction are magnetic susceptibility measurements. At room temperature an effective magnetic moment $\mu_{eff}$ = 4.9 $\mu_B$ is expected for an iron(II) HS isomer, whereas $\mu_{eff}$ should vanish for a pure LS isomer. Due to spin-orbit coupling the measured $\mu_{eff}$ for HS isomers is usually larger than 5 $\mu_B$, and in case of near degeneracy of the electronic ground state even effective magnetic moments larger than 6 $\mu_B$ have been measured [7,8]. From the temperature-dependent effective magnetic moment the molar HS fraction has been derived as a function of temperature (Figure 3). The resulting transition temperatures are in the range between 175 and 355 K (Table 2). For the compounds **2**[a] and **5**[b] the transition temperature determined by magnetic susceptibility measurements is close to zero and it might be that these complexes undergo only a partial or no spin-transition. Except for compound **4**[b] which has roughly the same transition temperature as compound **1**[a], all other compounds exhibit a lower transition temperature. Substituting a hydrogen atom of the pyrazol ring by a methyl group, like in compounds **2**[a], **2**[b], **2**[c], and **3**[a], decreases $T_{1/2}$. The decrease is even stronger if two hydrogens are substituted as is the case for compound **5**[b]. Comparison of compounds **2**[a], **2**[b], and **2**[c] reveals a significant influence of the counteranion on $T_{1/2}$.

**Table 2.** $T_{1/2}$ in Kelvin as derived from magnetic susceptibility measurements (Figure 3). The theoretical values were calculated with BLYP//LANL2DZ and a constant was added afterwards in order to match the experimental value for compound **1**[a].

| Complex | Experiment | Theory |
|---|---|---|
| **1**[a] | ≈ 355[a] | 355[d] |
| **2**[a], **2**[b], **2**[c] | ≈ 0[a], 175[b], 220[c] | < 0[e] |
| **3**[a] | ≈ 430[a] | |
| **4**[b] | ≈ 355[b] | 335[e] |
| **5**[b] | ≈ 0[b] | |

[a] with counterion $PF_6^-$, [b] with counterion $ClO_4^-$, [c] with counterion $BF_4^-$, [d] by definition, [e] see also eq. (4) and text below

*Density Functional Calculations*

From the DFT calculations for complexes **1** to **5** the difference of electronic energy $\Delta E_{el}$, of vibrational energy $\Delta E_{vib}$, and of entropy $\Delta S$ can be retrieved. It has been assumed here, that the lowest electronic excitation energy is large compared to $k_B T$, and hence $\Delta E_{el}$ is regarded to be constant in the temperature range of interest. The term $p\Delta V$ can not be calculated in the molecular approximation applied here. A typical value of $\Delta V \approx 7$ cm$^3$ mol$^{-1}$ [1] leads to a contribution to the free energy of $p\Delta V \approx 0.7$ J mol$^{-1}$ at ambient pressure and temperature. This is far less than the error margin of the other contributions to $\Delta G$, and the neglect of $p\Delta V$ is therefore justified.

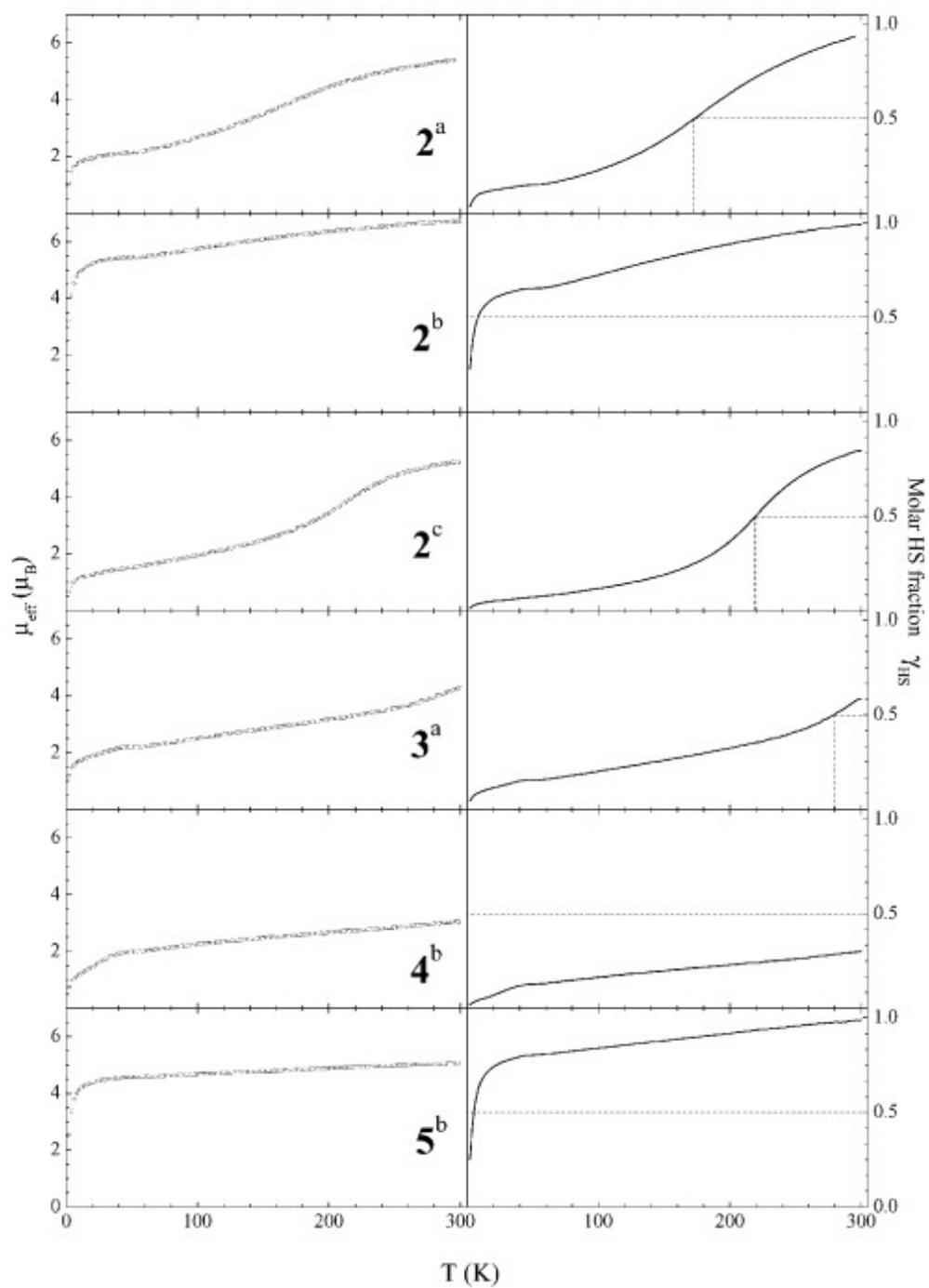

**Figure 3.** Temperature dependent effective magnetic moment and derived molar HS fraction of compounds as indicated.

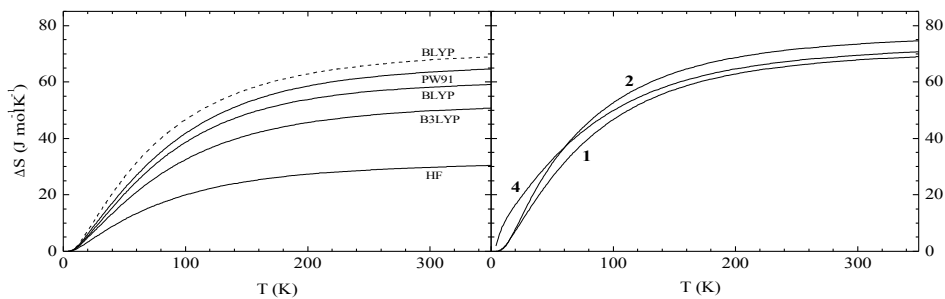

**Figure 4.** $\Delta S$ for complex **1** as a function of temperature calculated with different methods as indicated (left panel, solid lines refer to the 6-311G basis, the dashed line refers to the LANL2DZ basis) and for complexes **1**, **2**, and **4** calculated with BLYP//LANL2DZ (right panel).

All calculated terms of the free energy depend significantly on the chosen method and basis set. For the temperature-dependent entropy difference of complex **1** the largest observed deviations are between the pure DFT methods like BLYP or PW91 on one side and the Hartree-Fock method on the other side (Figure 4). The calculated curves $\Delta S(T)$ for the other compounds are affected in a similar way by the choice of the computational method. The same is valid for the temperature-dependent vibrational energy difference $\Delta E_{vib}(T)$. The curves calculated for complexes **1**, **2**, and **4** using the BLYP//LANL2DZ method are almost identical (Figure 5).

The largest contributions to the free energy arise from the electronic energy difference $\Delta E_{el}$ (Table 3). The Hartree-Fock method gives negative energy differences and thus fails to predict the correct LS ground state at low temperature. To a lesser extent this is also true for the hybrid method B3LYP which includes about 20 % HF contribution to the exchange energy. The reason for this behaviour is that the HF methods, which per definition does not include any electron correlation, favours higher spin-states where the Pauli principle forces the electrons to avoid each other [9,10]. It is interesting to note that the calculated shifts of $\Delta E_{el}$, when going from the unsubstituted complex **1** to the substituted complexes, seem to be of similar order of magnitude for the HF methods as for the DFT methods (Table 3). From the computational methods used in this study the pure DFT methods BLYP and PW91 seem to be the most reliable ones for the calculation of the free energy difference, since these methods always give the correct LS groundstate.

**Table 3.** $\Delta E_{el}$ in kJ mol$^{-1}$ calculated with 6-311G and LANL2DZ basis sets and with the Hartree-Fock and different DFT methods (values in brackets give the difference to complex **1**).

| Complex | HF | | B3LYP | | BLYP | | PW91 | |
|---|---|---|---|---|---|---|---|---|
| | 6-311G | LANL2DZ | 6-311G | LANL2DZ | 6-311G | LANL2DZ | 6-311G | LANL2DZ |
| **1** | -300.158 | -280.561 | -7.302 | 9.607 | 81.721 | 84.953 | 110.768 | 103.552 |
| **2** | | -301.276 | -39.511 | -23.853 | 39.117 | 43.253 | | 60.423 |
| | | (20.715) | (32.210) | (33.459) | (42.604) | (41.701) | | (43.129) |
| **3** | | -280.779 | | | | | | |
| | | (-0.218) | | | | | | |
| **4** | | -280.385 | | | 79.116 | 83.252 | | |
| | | (-0.176) | | | (-2.605) | (-1.701) | | |

Unfortunately the term $\Delta E_{el}$, which dominates the free energy $G$, is the term with the largest error margin. The reason is that the difference $\Delta E_{el}$ is more than five orders of magnitude smaller than the absolute values of $E_{el}$. Qualitatively correct values of $\Delta E_{el}$ can, therefore, only be obtained if the systematic errors inherent in the individually calculated values $E_{el}$ for HS and LS states cancel when the difference $\Delta E_{el}$ is formed. For $\Delta E_{vib}$ and $\Delta S$ the relation between the differences and the absolute values is by more than three orders of magnitude more favourable than for $\Delta E_{el}$ and the absolute values $E_{el}$ for HS and LS states, and it has been demonstrated in the past that the normal modes of molecular vibrations, from which $\Delta E_{vib}$ and $\Delta S$ are derived, can be calculated for such complexes with reasonable accuracy [4,5,6].

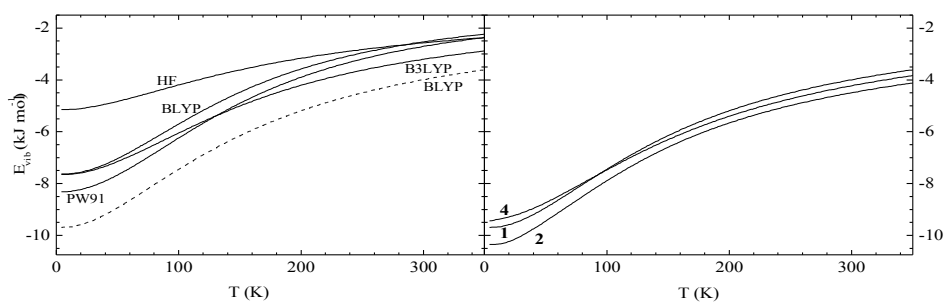

**Figure 5.** $\Delta E_{vib}$ for complex **1** calculated with different methods as indicated (left panel, solid lines refer to 6-311G basis, dashed line refers to LANL2DZ basis) and for complexes **1**, **2**, and **4** calculated with BLYP//LANL2DZ (right panel).

The temperature-dependent free energy difference $\Delta G(T)$ (Figure 6) is calculated by summing up the terms given in eq. (2) except for the volume change which is neglected. Ideally, the points of intersection of the curves $\Delta G(T)$ with the abscissa should yield the transition temperatures $T_{1/2}$, according to its definition in eq. (1). However, it is obvious that the calculated curves are shifted due to errors of $\Delta E_{el}$, which prevents the determination of absolute values for $T_{1/2}$. A rough estimate for the difference $\Delta T_{1/2}$ of transition temperatures when comparing two complexes **a** and **b** can be obtained by the expression

$$\Delta T_{1/2} \approx (\Delta E_{el}^{b} - \Delta E_{el}^{a}) / \Delta S(T_{1/2}^{a}), \qquad (4)$$

where $T_{1/2}^{a}$ must be known from experiment and $\Delta S^{a}(T) \approx \Delta S^{b}(T)$ is assumed. Comparison of the shift of the transition temperature estimated in this way with experimental values (Table 2) yields agreement for the direction of the shift and for the order of magnitude of the shift.

*Conclusions*

It has been shown for the example of bis-tripodal chelates of iron(II) with tris(pyrazolyl)methane ligands that the temperature of spin transition can be influenced significantly if hydrogen atoms of the pyrazol rings are substituted by methyl groups. These ligands are promising materials for the finetuning of technically interesting properties of spin-crossover complexes. Because of the large variety of possible substitutions it would be very helpful if the effect of these substitutions could be tested by computer simulations with at least qualitative accuracy. The calculations presented here suggest that it should be possible with currently available density functional methods to predict the direction and the order of magnitude of a shift of the transition temperature. The dominating source of error is the calculation of the electronic energy difference.

Future calculations with periodic boundary conditions will be performed to study the influence of intermolecular interactions and of counterions.

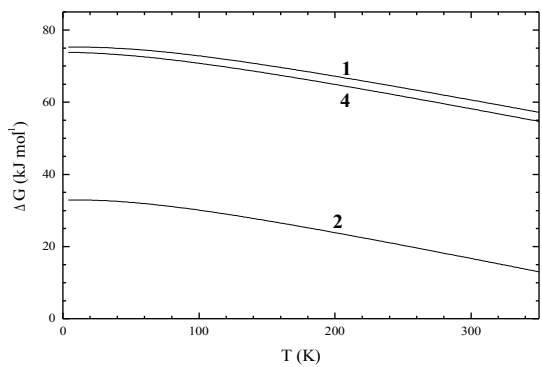

**Figure 6.** $\Delta G$ for complexes **1**, **2**, and **4** calculated with BLYP//LANL2DZ.

## Materials and Methods

*Synthesis*

The ligands were synthesized by the method described in Ref. [11,12,13], which consists of reacting the substituted pyrazol with chloroform under solid liquid phase transfer conditions. The ligands where synthesised by refluxing the substituted pyrazol, finely powdered potassium carbonate and tetrakis(n-butyl)ammonium bromide in chloroform for approximately 30 hours. After filtering of the solid potassium carbonate the solution was purified by different methods. Tris(3,5-dimethylpyrazol-1-yl)methane was purified by stirring the solution with charcoal, washing with hexane and after removing the solvent sublimating (420 K, 0.1 Torr). Tris(3-methylpyrazol-1-yl)methane, tris(4-methyl-pyrazol-1-yl)methane, and tris(4-bromo-pyrazol-1-yl)methane were purified by removing the chloroform under reduced pressure and crystallization from diethyl ether at 5 °C.

$^1$H-NMR (CDCl$_3$,ppm relative to TMS): tris(3,5-dimethylpyrazol-1-yl)methane: 8.07, 5.87, 2.18, 2.01, tris(3-methylpyrazol-1-yl)methane: 8.21, 7.55, 7.32, 6.11, 2.40, 2.28, tris(4-methyl-pyrazol-1-yl)methane: 8.17, 7.48, 7.32, 2,08, tris(4-bromo-pyrazol-1-yl)methane: 8.22, 7.65, 7.27, 2.17.

$^{13}$C-NMR (CDCl$_3$,ppm) tris(3,5-dimethylpyrazol-1-yl)methane: 148.77, 140.77, 107.62, 80.77, 13.82, 10.76, tris(3-methylpyrazol-1-yl)methane: 141.11, 120.73, 107.05, 80.75, 14.15, 10.75, tris(4-methyl-pyrazol-1-yl)methane: 142.42, 127.71, 117.64, 83.21, 8.85, tris(4-bromo-pyrazol-1-yl)methane: 142.9, 129.71, 96.24, 83.84.

The iron(II) complexes were synthesized in ethanol by mixing stochiometric amounts of the ligand and either Fe(ClO$_4$)$_2$ for the perchlorate salt, or Fe(SO$_4$)$_2$·7H$_2$O for the other salts. The appropriate anion was then added and the complex was obtained as a powder, which was washed with ethanol and water. For all complexes satisfactory elementary analysis where obtained.

*Spectroscopy*

Conventional Mössbauer spectra were obtained in transmission geometry. The $^{57}$Co[Rh] source was driven with constant acceleration. The energy calibration was performed with α-iron at room temperature and the isomer shift is relative to this standard.

For the magnetic measurements a SQUID magnetometer of the type MPMS Quantum Design was used. The applied field was 1 T for all temperatures. The magnetic susceptibility measurements were performed at various temperatures between 2 K and 300 K by decreasing and increasing the temperature in steps of 2 K. The effective magnetic moment μ$_{eff}$ expressed in units of the Bohr magneton μ$_B$ was derived from the experimental data by

$$\mu_{eff} = \mu_B \sqrt{8T(w_{mol}\chi_{mass} - \chi_D)}. \qquad (5)$$

In this expression $\chi_D$ denotes the molar diamagnetic susceptibility which was estimated from Pascals constants [14], $w_{mol}$ is the molecular weight of the compounds and $\chi_{mass}$ the experimental mass susceptibility at temperature $T$.

NMR spectra where recorded on a 300 MHz Bruker AC 300.

*Density Functional Calculations*

Several different density funtional methods were used: (i) Perdew and Wang's exchange functional and their gradient corrected correlation functional [15] (PW91), (ii) Becke's exchange functional [16] together with Perdew's gradient corrected correlation functional [17] (BP86), (iii) Becke's exchange functional [16] using the correlation functional of Lee, Yang, and Parr [18,19] (BLYP), and (iv) Becke's three parameter hybrid functional [20] using Lee, Yang, and Parr's correlation functional [18,19] (B3LYP). We used the following basis sets: (i) the 6-311+G(2d,p) basis for H, C, and N and the Wachters-Hay double-ζ basis for Fe [21,22] (6-311 for short), (ii) the Dunning-Huzinaga all-electron double-ζ basis for H, C, and N and the Los Alamos effective core potential plus double-ζ basis set on Fe [23,24] (LANL2DZ). The calculations for the anion were performed with the program packages Gaussian 98 [25] and Turbomole [26].

All calculations were performed for molecules *in vacuo*. The total energy $E_{el}$ for HS and LS states was calculated after full geometry optimization for the respective spin states. For part of the complexes the vibrational frequencies were calculated in order to determine $E_{vib}(T)$ and $S(T)$

$$E_{vib}(T) = k_B T \sum_i x_i \coth(x_i) \qquad (6)$$

according to the relations
and
using the definition $x_i = \hbar\omega_i/2k_B T$ and denoting the angular frequency of the *i*th vibrational normal

$$S(T) = k_B \sum_i \left\{ x_i \coth(x_i) - \ln[2\sinh(x_i)] \right\} \qquad (7)$$

mode by $\omega_i$.

## Acknowledgements

Financial support by the Deutsche Forschungsgemeinschaft (DFG) within the priority programme *Molecular Magnetism* is gratefully acknowledged. MEMPHYS – Center for Biomembrane Physics is supported by the Danish National Research Foundation.